\documentclass[aps,pra,twocolumn,superscriptaddress,showpacs]{revtex4-1}
\usepackage{graphics}
\usepackage{graphicx}
\ifx\pdftexversion\undefined
\usepackage[dvips]{hyperref}
\else
\usepackage{hyperref}
\fi
\hypersetup{
  colorlinks = true, linkcolor = blue
}

\begin{document}
\preprint{}

\title{Fast production of Bose-Einstein condensates of metastable Helium}

\author{Q. Bouton} 
\author{R. Chang} 
\author{A. L. Hoendervanger}
\author{F. Nogrette}
\author{A. Aspect}
\author{C. I. Westbrook}
\author{D. Cl\'ement}
\affiliation{Laboratoire Charles Fabry, Institut d'Optique, CNRS, Univ. Paris Sud, 2 Avenue Augustin Fresnel 91127 PALAISEAU cedex, France}

\date{\today}

\begin{abstract}
We report on the Bose-Einstein condensation of metastable Helium-4 atoms using a hybrid approach, consisting of a magnetic quadrupole and a crossed optical dipole trap. In our setup we cross the phase transition with $2\times10^{6}$ atoms, and we obtain pure condensates of $5\times10^{5}$ atoms in the optical trap. This novel approach to cooling Helium-4 provides enhanced cycle stability, large optical access to the atoms and results in production of a condensate every 6 seconds -- a factor 3 faster than the state-of-the-art.  This speed-up will dramatically reduce the data acquisition time needed for the measurement of many particle correlations, made possible by the ability of metastable Helium to be detected individually. 
\end{abstract}

\maketitle 



In the study of many physical phenomena,  one often seeks to make highly accurate measurements or search for rare events. In these situations convincing evidence can only be obtained by acquiring a large amount of data to increase the statistical significance of the results. Multi-particle correlations are a good example of such measurements and are often used in both particle physics \cite{CMS2010} and quantum optics \cite{Mandel1995, Grynberg2010}. 
Thus efforts to increase data acquisition rates often constitute a significant preoccupation for experimentalists. 

In the field of ultracold gases recent advances in imaging techniques have permitted the detection of individual atoms and the measurement of their correlations \cite{Schellekens2005, Nelson2007, Gericke2008, Bucker2009, Bakr2009, Sherson2010, Haller2015, Cheuk2015}. Because of the electronic detection techniques enabled by the use of metastable Helium He$^*$, significant contributions to this field have been made using this species \cite{Vassen2012, Dall2013, Lopes2014, Lopes2015}. However these experiments are often limited by the stability of the apparatus over long data runs. 

For many atomic species, the stability and cycle time have been dramatically improved by all-optical cooling \cite{Barret2001, Clement2009} or by magnetic-optical hybrid traps \cite{Lin2009}. Such techniques have yet to be realized in He$^*$, in part because of significant challenges posed by the unusual characteristics of this atom. In unpolarized He$^*$, the density is severely limited by Penning collisions \cite{Bardou1992, Mastwijk1998, Kumakura1999, Browaeys2000} preventing the use of all-optical cooling techniques without spin polarization. Hybrid traps use magnetic quadrupoles in which one must minimize losses due to Majorana spin flips. Since the Majorana loss rate, at a given temperature, varies inversely with the particle mass \cite{Majorana1932}, He* is particularly unfavorable for this approach as well. Nevertheless the potential gain offered by these novel approaches is sufficiently great that it is worth some effort to try to circumvent these difficulties. 

\begin{figure}[h!]
\centering
\includegraphics[width=1\columnwidth]{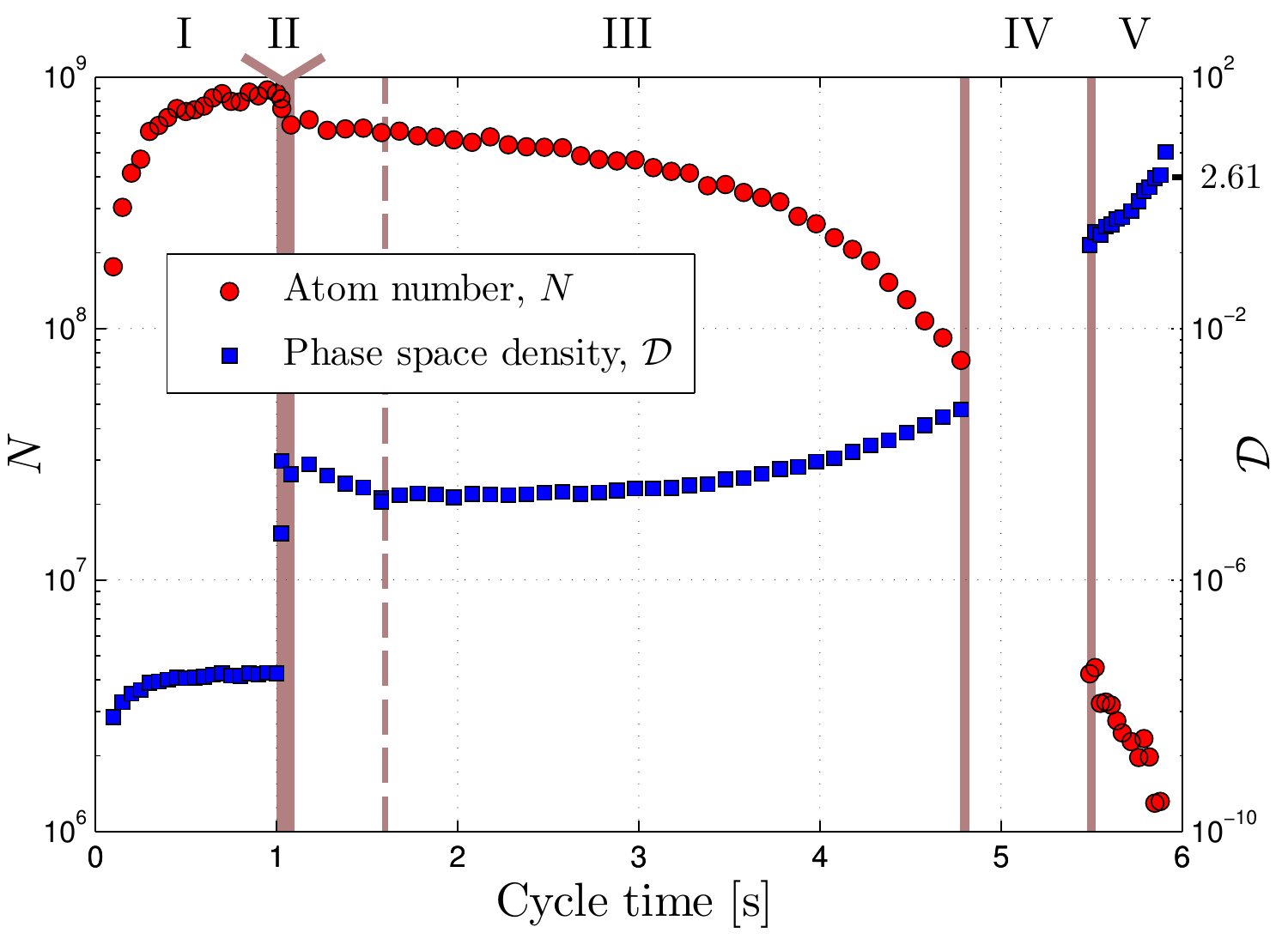}
\caption{(Color online).  Atom number $N$ (red circles) and phase space density $\mathcal{D}$ (blue squares) through the  experimental cycle.  The scheme is divided in stages: (I) Magneto-optical trap loading and compression, (II) Doppler and gray molasses cooling, (III) compression and radio-frequency evaporation in a quadrupole magnetic trap (QMT), (IV) transfer from the QMT to the optical dipole trap (ODT), and (V) forced evaporation in the ODT to Bose-Einstein condensation.}
\label{Fig:RampData}
\end{figure}

Here we report the implementation of a hybrid approach to Bose-Einstein condensation (BEC) of Helium in the metastable state ($^4$He$^*$), permitting fast and robust production of condensates with $5 \times 10^5$ atoms every 6 seconds. We demonstrate that the usage of a quadrupole magnetic trap (QMT) in place of the Ioffe or cloverleaf configurations used in other $^4$He$^*$ experiments \cite{Robert2001, Santos2001, Tychkov2006, Dall2006, Doret2009,Keller2014} offers technical simplicity, enhanced stability and shorter cycle time -- a factor 3 speed-up over the state-of-the-art.  Starting from a magneto-optical trap, our experimental cycle consists of laser cooling in optical molasses (Doppler and gray molasses), compression and radio-frequency evaporation in a QMT, transfer into a crossed optical dipole trap (ODT) by opening the QMT, and final forced evaporation to BEC. The cloud properties throughout the cooling cycle are summarized in Table~\ref{Table:Summary}, and the atom number $N$ and phase-space density $\mathcal{D}$ are plotted in Fig.~\ref{Fig:RampData}. As it can be seen from Table~\ref{Table:Summary}, each step of the cycle gives a substantial contribution to the fast production of BECs.

\begin{table*}[ht]
\begin{ruledtabular}
\begin{tabular}{ccccc}
Experimental steps &  $N$ & $T$ [$\mu$K] & $\rho_0$ [$\mathrm{cm^{-3}}$] & $\mathcal{D}$  \\
\hline
Magneto-Optical Trap & $8.7(2) \times 10^{8}$ & 1200(100) & $2.5(1) \times 10^{9}$& $4.0(4) \times 10^{-8}$\\
Red Molasses & $8.2(1) \times 10^{8}$ & 120(11) & $1.1(1) \times 10^{10}$ &  $5.5(6) \times 10^{-6}$\\
Gray Molasses & $7.5(2) \times 10^{8}$ & 20(3) & $1.1(1) \times 10^{10}$ &  $7.8(7) \times 10^{-5}$\\
Quadrupole Trap loading & $6.4(2) \times 10^{8}$ & 38(3) & $2.0(2) \times 10^{10}$ & $5.5(2) \times 10^{-5}$ \\
Quadrupole Trap compression & $6.1(1) \times 10^{8}$ & 230(15) & $1.0 (1)\times 10^{11}$ & $ 2.0(3) \times 10^{-5}$ \\
Radio-frequency evaporation & $6.0(1) \times 10^{7}$ & 70(6) & $3.3(6)\times 10^{11}$ & $3.7(7) \times 10^{-4} $\\
Optical Dipole Trap loading & $5.0(2) \times 10^{6}$ & 27(2)  & $5.1(2) \times 10^{13}$ & $2.4(3) \times 10^{-1} $ \\
\end{tabular}
\caption{Measured cloud parameters following each phase of the cooling cycle.  We present the atom number $N$, temperature $T$, peak density $\rho_0$ and phase space density $\mathcal{D}$.} \label{Table:Summary}
\end{ruledtabular}
\end{table*}

\emph{Magneto-Optical Trapping and Laser Cooling}.  The first stage has already been described in detail in \cite{Chang2014}.  Briefly, a six-beam MOT of $^4$He$^*$ atoms is loaded from a dc-discharge plasma source, followed by transverse (molasses) and longitudinal (Zeeman slower) cooling.  The MOT is composed of a quadrupole magnetic field with gradient $B^\prime=25$ G/cm (strong-axis) and light red-detuned by $\Delta=-31 \Gamma$ ($\mathrm{\Gamma/2\pi = 1.6}$  MHz) from the $2^3$S$_1\rightarrow 2^3$P$_2$ resonance and at a high intensity of 50 $\mathrm{I_{sat}}$ per beam ($\mathrm{I_{sat}=0.16 \ mW/cm^{2}}$).  This configuration provides a large capture velocity while keeping the cloud density low so to reduce losses due to light-assisted Penning collisions \cite{Bardou1992, Mastwijk1998, Kumakura1999}. MOTs with $N \sim 9 \times 10^{8}$ atoms at a temperature of $T=1.2$ mK are loaded in 1 second \cite{Fluo}.  The MOT is then compressed by ramping the detuning $\Delta$ from $-31 \Gamma$ to $-2 \Gamma$ in 20 ms.  At the same time, the light intensity is reduced from $\mathrm{50 \ I_{sat}}$ to $\mathrm{0.33 \ I_{sat}}$ to reduce the rate of light-induced Penning collisions. At the end of this phase, we have increased the density by a factor $\simeq 9$ while loosing only a small fraction of atoms (10 \%). 

After the compression we further cool the cloud in optical molasses. We first use a red molasses (RM) to capture the atoms from the MOT and then load them into a gray molasses (GM) to reach low temperatures. Although it is possible to cool $^4$He$^*$ on the $\mathrm{2^{3}S_{1} \ \rightarrow \ 2^{3}P_{2}}$ transition to the Doppler limit ($T_D=\hbar\Gamma/2k_B=38$ $\mu$K) \cite{Chang2014}, we cool only to $\mathrm{120 \ \mu K}$ in 5 ms, a value that is sufficiently low to load a blue-detuned gray molasses on the $\mathrm{2^{3}S_{1} \ \rightarrow \ 2^{3}P_{1}}$ transition. Three-dimensional cooling of Helium gases with blue-detuned molasses has been demonstrated in the past \cite{Lawal1995}. While sub-Recoil cooling was achieved, only a small fraction of $10^4$ atoms were brought to these ultra-low temperatures. Here the parameters we choose (laser detuning $\Delta'=+10\Gamma$ and intensity of $20$ $\text{I}_{\text{sat}}$) allow us to capture nearly all the atoms from the RM and cool $N=7.5 \times 10^{8}$ atoms to a temperature of $T=20$ $\mu$K in 5 ms. At this stage the low atomic density $\rho_{0} \simeq 1.1 \times 10^{10}$ cm$^{-3}$, limited by light-induced Penning collisions in the MOT, prevents us from efficiently loading an optical dipole trap. On the other hand, a spin-polarized Helium gas in the absence of near-resonant light has a significantly smaller cross-section for collision \cite{Shlyapnikov1994}. Thus we load the atoms into a magnetic trap where the atomic density can be increased in order to facilitate efficient transfer into the optical trap.


\emph{QMT loading and compression.} We first optically pump a large fraction of the atoms ($\sim 85 \%$) to the $J=1$ $m_{J}=1$ state. The gas is then captured by abruptly turning on a magnetic field gradient of $B^\prime=4.5$ G/cm. The magnetic field is produced by the pair of coils in anti-Helmholtz coils used for the MOT. Next we adiabatically ramp up the field to $B^\prime=45$ G/cm over 500 ms in order to increase the density and thus the elastic collision rate.  At this point there are $N=6 \times 10^8$ atoms in the compressed cloud at a temperature of $T=230$ $\mu$K, corresponding to a phase space density of $\mathcal{D}=2.0 \times 10^{-5}$.  

\begin{figure}
\centering
\includegraphics[width=1\columnwidth]{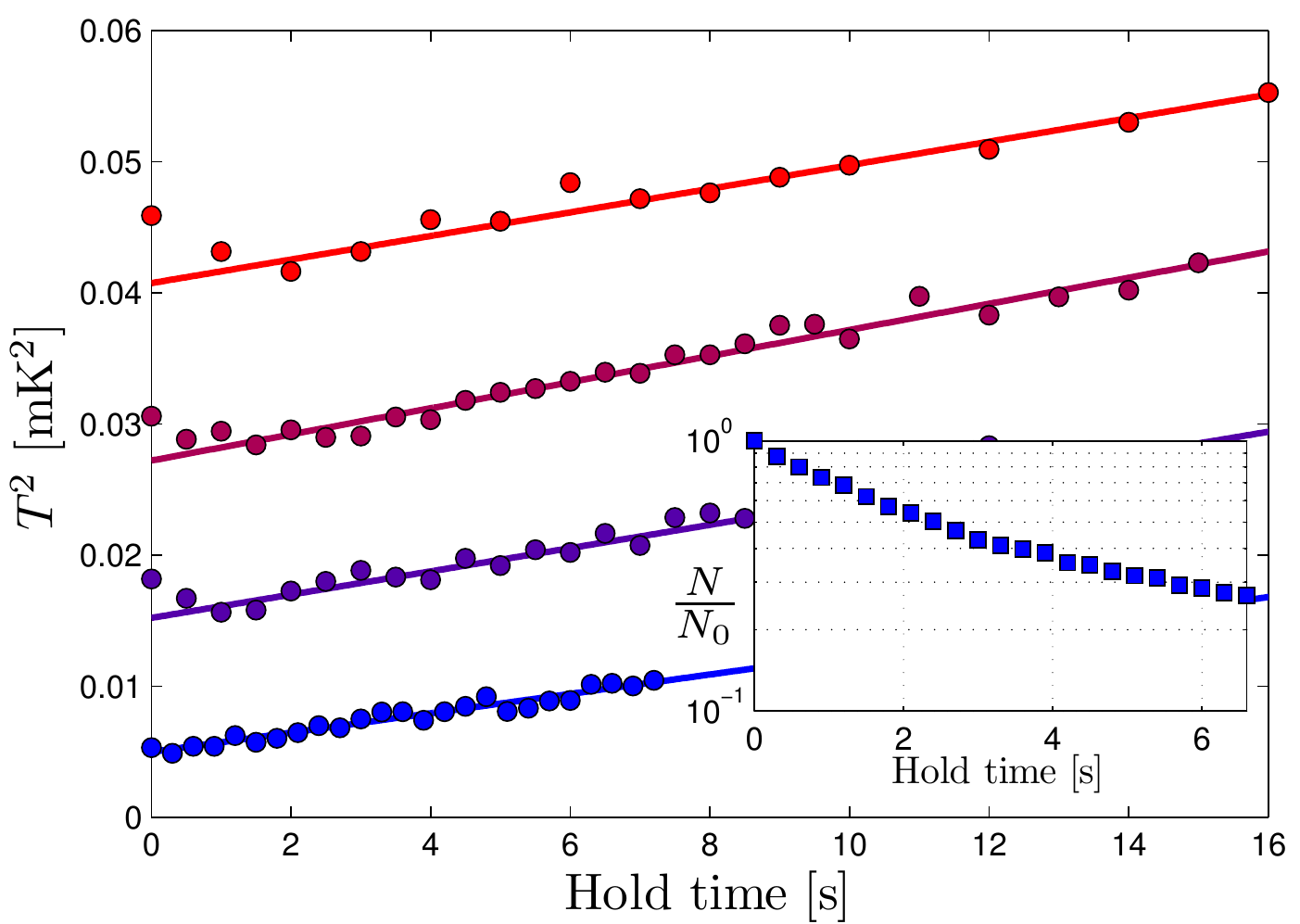}
\caption{Time evolution of the square of the cloud temperature $T$ in a QMT with fixed $B^\prime=45$ G/cm and for varying initial temperatures $T_0=202, 165, 123$ and 70 $\mu$K (dots).  Lines are fits according to Eq.~\ref{Eq:Majorana_T} from which we extract a geometrical factor $\chi=0.17(2)$. Inset: Evolution of the atom number in the QMT for the data set with $T_0=$ 70 $\mu$K.}
\label{Fig:Majorana}
\end{figure}

A well known issue with the QMT is the existence of a magnetic field zero at the center of the trap, which can lead to Majorana spin-flips to an un-trapped magnetic sub-state \cite{Majorana1932}.  At a given temperature the rate of spin-flips scales inversely with the particle mass and thus is particularly strong for Helium. The effects of Majorana spin-flips are two-fold: firstly they result in loss of atoms, as atoms escape from the $m_J=1$ trapped state to the un-trapped states ($m_J=0$ and $m_J=-1$).  The Majorana loss rate is \cite{Heo2011,Petrich1995} 
\begin{equation}
\Gamma_{m}=  \chi \frac{\hbar}{m} \left(\frac{\mu B'}{k_{B}T}\right)^{2},
\label{Eq:Majorana_Gm}
\end{equation}
where $k_{B}$ is the Boltzmann constant, $B'$ the magnetic gradient along the coil axis, $\mu=2 \mu_{B}$ the magnetic moment, $\mu_{B}$ the Bohr magneton, $\hbar$ Planck's constant, $m$ the particle mass and $\chi$ a dimensionless geometrical factor.  Secondly, this loss results in a heating of the cloud, as it is typically the coldest atoms, that stay around the field zero, that are lost. This change in temperature results in a non-exponential decay of the atom number $N$ (see the inset of Fig.~\ref{Fig:Majorana}). The temperature evolution is expected to follow a simple law \cite{Dubessy2012}
\begin{equation}
T(t)=\sqrt{T_{0}^{2} + \gamma_{m}t},
\label{Eq:Majorana_T}
\end{equation}
with 
\begin{equation}
\gamma_{m}=\frac{8}{9} \chi \frac{\hbar}{m} \left(\frac{\mu B^{'}}{k_{B}}\right)^{2}
\label{Eq:Majorana_gm}
\end{equation}
\noindent
where $T_{0}$ is the initial temperature at $t=0$. 

We monitor the cloud temperature as a function of time in a fixed magnetic trap and for different initial temperatures $T_{0}$. Our measurements, shown on Fig~\ref{Fig:Majorana}, are in excellent agreement with this prediction, with a fitted geometrical factor $\chi=0.17(2)$. This value agrees well with that reported for other species \cite{Heo2011, Dubessy2012}.

The measurement of the heating and atom loss rates allows us to tailor our compression and cooling cycle to minimize these negative effects.  To this end, we compress the magnetic trap to a final gradient of only 45 G/cm. Even at this relatively weak gradient the density is increased by a factor 30 (see Table~\ref{Table:Summary}) and the initial elastic collision rate of 170 s$^{-1}$ (the vacuum lifetime is 50 sec) is high enough to perform forced evaporation \cite{Browaeys2001}. 


\emph{Radio-frequency evaporation in the QMT.} Cooling in the QMT is performed by forced radio-frequency (RF) evaporation.  The RF signal is generated by a National Instruments PXI-5406A arbitrary function generator and amplified to 50 W by a Prana DP140 RF amplifier.  The field is generated by an RF coil outside the vacuum chamber, positioned 6 cm from the atoms.  An RF field of frequency $\nu$ imposes a trap depth which is characterized
by the cutting parameter $\eta=h\nu/k_BT$ \cite{Ketterle1996}.  Our ramp begins with a constant frequency of 34 MHz for 1 second followed by a linear sweep of the frequency from 34 MHz to 8 MHz during 2.25 seconds.  The RF cooling is summarized in stage III of Figure~\ref{Fig:RampData} where the vertical dashed line marks the beginning of the RF evaporation.  During this stage, the peak collision rate increases from 170 s$^{-1}$ to 360 s$^{-1}$, while the Majorana loss rate $\Gamma_m$ increases from $8.6 \times 10^{-3}$  s$^{-1}$ to $1.0 \times 10^{-1}$ s$^{-1}$.  The final cloud has $N=6 \times 10^{7}$ atoms at a temperature of $T=70$ $\mu$K, with a peak atomic density $\rho_{0}=3.3 \times 10^{11}$ cm$^{-3}$ below the densities at which inelastic 2-body and 3-body loss become important \cite{Partridge2010, Santos2001, Tychkov2006, Dall2006}. Having gained more than one order of magnitude in the density $\rho_{0}$, we begin the transfer into the optical trap. 



\emph{Transfer to the optical dipole trap and Bose-Einstein condensation.} The optical dipole trap is made of two gaussian-shaped beams at 1550 nm (fiber laser source IPG ELR-30-1550-LP).  The first beam has a power of 18 W, focused down to a  waist of 133 $\mu$m. The second beam contains 8 W focused to a waist of 63 $\mu$m, and crosses the first beam at an angle of $40^{\circ}$ in the horizontal plane.  The resulting trap is roughly cylindrical with trap frequencies of 3.1 kHz radially and 624 Hz axially, and a trap depth of $U_0=k_{B} \times$ 244 $\mu$K.  The trap center is displaced roughly one beam waist below the quadrupole center to avoid accumulating atoms at the magnetic field-zero.


The transfer from the QMT to the ODT is performed by ramping down the field gradient to 3 G/cm over 600 ms in the presence of the ODT (stage IV in Fig.~\ref{Fig:RampData}), before abruptly switching it off. A small bias field maintains the spin-polarization of the atoms in the ODT. Due to the overlap of magnetically trapped and optically trapped atoms during this stage, images of the atom clouds are difficult to interpret quantitatively, hence it is only the start and end points which are reported in Fig.~\ref{Fig:RampData}.  We estimate there are up to $1.5\times 10^7$ atoms at a peak density of $\rho_0\simeq10^{14}$ atoms/cm$^3$ initially in the crossed ODT.  The atom number proceeds to decay rapidly due to a mixture of evaporative cooling and 3-body loss \cite{Santos2001, Tychkov2006, Dall2006}.  After 20 ms the atom cloud equilibrates with $N=5\times10^6$ atoms at a temperature of $T=25$ $\mu$K and a phase space density of $\mathcal{D}=0.2$. Despite retaining only 10\% of the atoms from the QMT, we observe a gain in phase space density of 3 orders of magnitude, similar to that reported in other hybrid trap setups \cite{Lin2009}. This increase comes from a combination of a dimple effect (resulting in a dramatic increase in the atomic density) and subsequent evaporation in the crossed ODT. 
Interestingly the final atom number loaded in the ODT does not strongly depend on the efficiency of the previous RF evaporation. Indeed the ODT atom number is similar for the 3.25 second RF ramp (overall efficiency of $\gamma=-d \log \mathcal{D}/d \log N=1.39$) as for a longer and more efficient 6.5 second ramp ($\gamma=1.62$). 

\begin{figure}[h!]
\centering
\includegraphics[width=1\columnwidth]{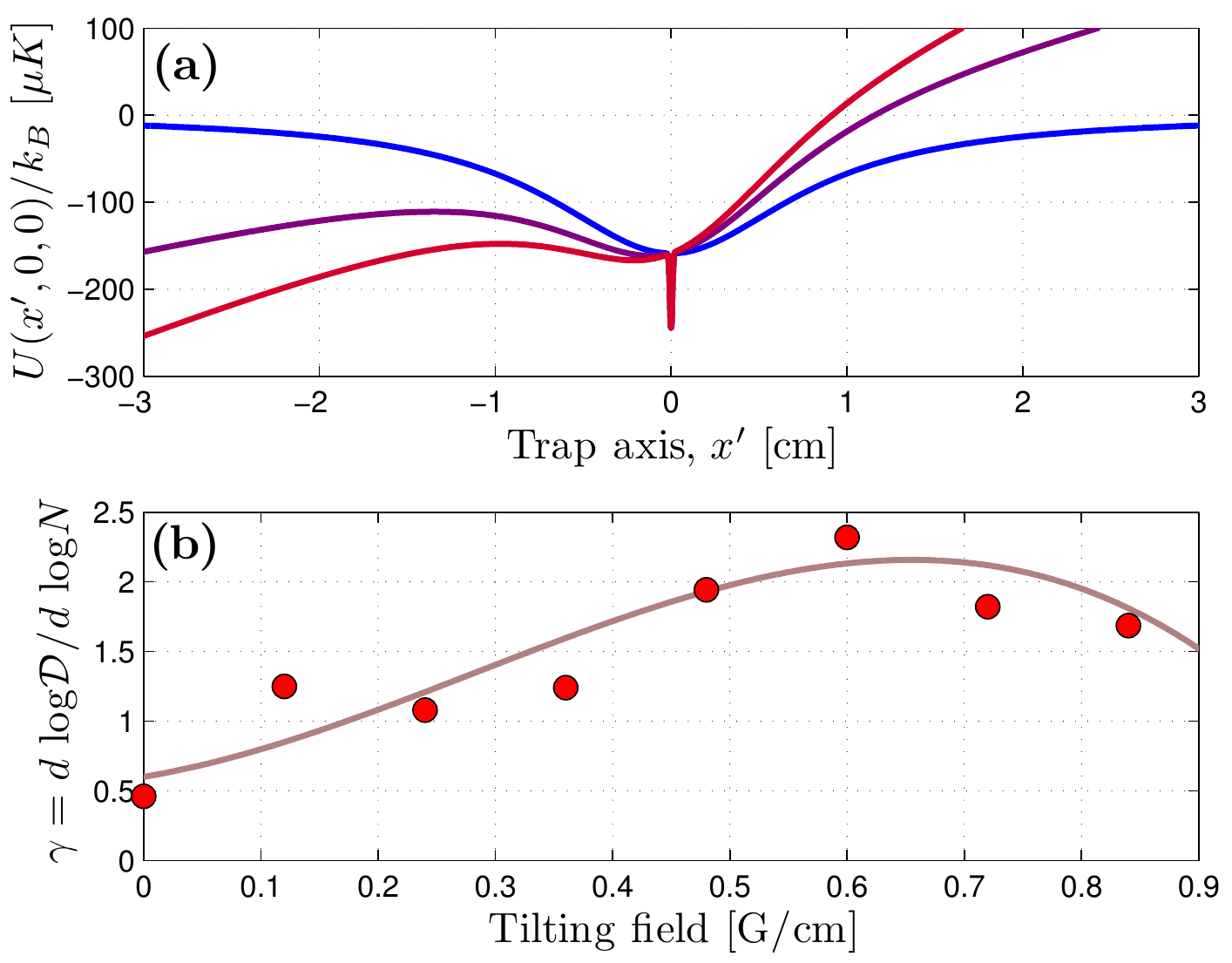}
\caption{(Color online) (a) ODT potential along the second beam axis and for tilting fields $B^\prime_{\text{tilt}}=0, 0.4$ and $0.6$ G/cm (blue, purple, red). (b) ODT evaporation efficiency $\gamma$ as a function of tilting field $B^\prime_{\text{tilt}}$. The solid line is a guide for the eye.}
\label{Fig:Tilting}
\end{figure}

The final cooling stage is performed by ramping down the ODT powers over 500 ms from 18 and 8 W to 1 and 0.3 W, in beams 1 and 2 respectively.  This evaporation is performed in the presence of a weak magnetic field gradient ($B^\prime_{\text{tilt}}=0.6$ G/cm) along the QMT coil axis $\hat{x}$.  Note that the resulting potential $U_{\text{tilt}}=\mu B^\prime_{\text{tilt}}x$ is weak relative to the crossed-optical potential ($U_0=k_{B} \times 244 \ \mu K$ initially), and thus does not serve as a means to change the trap depth as in other commonly used optical trap evaporation schemes \cite{Hung2008}.  Rather, the function of the tilting field is to remove the high-energy atoms which have been evaporated from the crossed-optical trap but remain in the individual optical trap beams (see Fig.~\ref{Fig:Tilting}).  

We observe that the evaporation efficiency rises from $\gamma=0.5$ in absence of tilt, to a maximum of $\gamma=2.3$ with a gradient of $B^\prime_{\text{tilt}}=0.46$ G/cm.  Above $B^\prime_{\text{tilt}}=0.6$ G/cm, the gradient begins to modify the ODT potential towards the end of the ramp and we observe a decrease in evaporation efficiency.  We note that the necessity of a gradient potential for ODT evaporation is particular to light atoms such as Helium, where the influence of gravity is feeble.  In heavier atoms such as $^{87}$Rb, a slight tilt in the optical beams of a few degrees with respect to gravity is often sufficient to remove these atoms. 

\begin{figure}[h!]
\centering
\includegraphics[width=1\columnwidth]{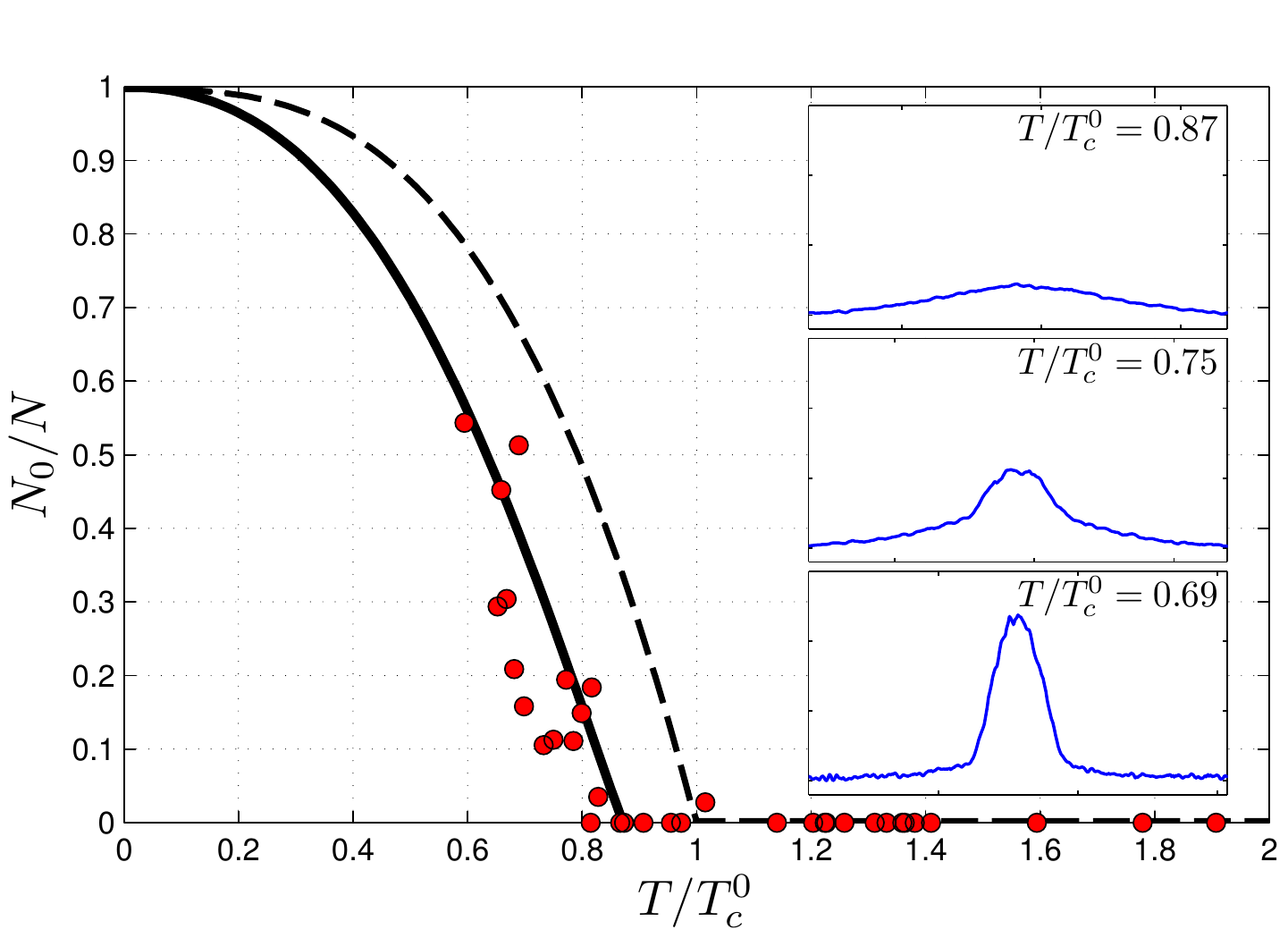}
\caption{(Color online). Condensate fraction across the BEC transition as a function of temperature.  The experimental data (red points) is compared to the theory for an ideal Bose gas (dashed black line) and to the theory including mean-field interaction in the BEC component (solid black line)  \cite{Dalfovo1999}. Data is plotted only for $T/T_c^0>0.6$ as fitting bi-modal distributions becomes unreliable at low temperatures. Inset: 1D Density profiles of the gas taken after time-of-flight and at various temperatures across the BEC transition.}
\label{Fig:BEC}
\end{figure}

With our favorable initial conditions after the transfer in the ODT ($N=5 \times 10^6$, $\mathcal{D}=0.2$ and an elastic collision rate of $3\times10^5$ s$^{-1}$), we reach the transition for Bose-Einstein condensation with $N=2\times10^6$ atoms at a temperature of $T=4$ $\mu$K. Figure~\ref{Fig:BEC} presents the measured condensate fraction as a function of temperature, normalized by the transition temperature $T_c^0$ of an ideal gas in a harmonic trap \cite{Dalfovo1999}. The condensate fraction and thermal gas temperature are estimated from a bi-modal fit of absorption images taken after time-of-flight. Finally we observe condensates of up to $N=5\times10^5$ atoms with no discernible thermal fraction. 


In conclusion, we have demonstrated a new approach to Bose-Einstein condensation of metastable Helium using a hybrid trap. The use of a magnetic quadrupole trap provides a means to reach high atomic densities allowing for efficient transfer into a crossed optical dipole trap, where fast and efficient forced evaporative cooling is performed. 
Several benefits result from this approach: (i) the 6 second ramp preparation time is a factor 3 shorter than the state-of-the-art, and will facilitate rapid measurements of correlation, permitted permitted by the ability of metastable Helium to be detected individually; (ii) the usage of a dipole trap for the evaporation to BEC in place of the Ioffe and cloverleaf trap configurations enhances the stability of the experimental cycle; (iii) the simplicity of the coil design (two coils in anti-Helmoltz configuration) provides a large optical access and thus an excellent numerical aperture to the atoms (N.A.=0.50). This paves the way to the investigation of many-particle correlations with metastable Helium atoms in experimental configurations of interest, for instance, engineered with optical potentials \cite{Bloch2008}. 

\begin{acknowledgments}
We acknowledge discussions with the members of the Atom Optics group at LCF and we thank Y. Fang, T. Klafka, A. Guilbaud, F. Moron and A. Villing for technical help.  We acknowledge financial support from the R\'egion Ile-de-France (DIM Daisy), the RTRA Triangle de la Physique, the European Research Council (Senior Grant Quantatop), the LabEx PALM (Grant number ANR-10-LABX-0039), the International Balzan Prize Foundation (2013 Prize for Quantum Information Processing and Communication awarded to A. Aspect), the Direction G\'en\'erale de l'Armement, and the Institut Francilien de Recherche sur les Atomes Froids.
\end{acknowledgments}


\begin{thebibliography}{21}
\expandafter\ifx\csname natexlab\endcsname\relax\def\natexlab#1{#1}\fi
\expandafter\ifx\csname bibnamefont\endcsname\relax
  \def\bibnamefont#1{#1}\fi
\expandafter\ifx\csname bibfnamefont\endcsname\relax
  \def\bibfnamefont#1{#1}\fi
\expandafter\ifx\csname citenamefont\endcsname\relax
  \def\citenamefont#1{#1}\fi
\expandafter\ifx\csname url\endcsname\relax
  \def\url#1{\texttt{#1}}\fi
\expandafter\ifx\csname urlprefix\endcsname\relax\def\urlprefix{URL }\fi
\providecommand{\bibinfo}[2]{#2}
\providecommand{\eprint}[2][]{\url{#2}}

\bibitem{CMS2010} CMS Collaboration, J. High Energy Phys. {\bf 09} 091 (2010).

\bibitem{Mandel1995} L. Mandel and E. Wolf,  {\it Optical Coherence and Quantum Optics}, Cambridge Univ. Press (1995).

\bibitem{Grynberg2010} G. Grynberg, A. Aspect, and C. Fabre, {\it Introduction to Quantum Optics: From the
Semiclassical Approach to Quantized Light}, Cambridge Univ. Press (2010)

\bibitem{Schellekens2005} M.Schellekens, R. Hoppeler, A. Perrin, J. Viana Gomes, D. Boiron, A. Aspect,  and C.I. Westbrook, Science {\bf 310} 648-651 (2005).

\bibitem{Nelson2007} K. D. Nelson, X. Li, and D. S. Weiss, D.S., Nature Phys. {\bf 3} 556 (2007).

\bibitem{Gericke2008} T. Gericke, P. Wurtz, D. Reitz, T. Langen and H. Ott, Nature Phys. {\bf 4} 949-953 (2008).

\bibitem{Bucker2009} R. Bucker, A. Perrin, S. Manz, T. Betz, C. Koller, T. Plisson, J. Rottmann, T. Schumm and J. Schmiedmayer, New J. Phys. {\bf 11} 103039 (2009).

\bibitem{Bakr2009} W. S.  Bakr,  J. I. Gillen,  A. Peng,  S. Folling,  and M. Greiner, Nature {\bf 461} 74-77 (2009).

\bibitem{Sherson2010} J. F. Sherson,   C. Weitenberg, M. Endres,   M. Cheneau, I. Bloch, and S. Kuhr, Nature {\bf 467} 68-72 (2010).

\bibitem{Haller2015} E. Haller, J. Hudson, A. Kelly, D. A. Cotta, B. Peaudecerf, G. D. Bruce, and S. Kuhr, arXiv:1503.02005 (2015).

\bibitem{Cheuk2015} L. W. Cheuk, M. A. Nichols,  M.   Okan, T.  Gersdorf,  V. V. Ramasesh, W. S. Bakr, T. Lompe and M. W. Zwierlein, arXiv:1503.02648 (2015).

\bibitem{Vassen2012} W. Vassen, C. Cohen-Tannoudji, M. Leduc, D. Boiron, C. I. Westbrook, A. Truscott, K. Baldwin, G. Birkl, P. Cancio and M. Trippenbach, Rev. Mod. Phys. {\bf 84} 175 (2012).



\bibitem{Dall2013} R. G. Dall, A. G. Manning, S. S. Hodgman, Wu RuGway, K. V. Kheruntsyan	and A. G. Truscott, Nat. Phys. {\bf 9}, 341–344 (2013).

\bibitem{Lopes2014} R. Lopes, A. Imanaliev, M. Bonneau, J. Ruaudel, M. Cheneau, D. Boiron, and C. I. Westbrook, Phys. Rev. A {\bf 90}, 013615 (2014).

\bibitem{Lopes2015} R. Lopes, A. Imanaliev, A. Aspect, M. Cheneau, D. Boiron, and C. I. Westbrook, Nature {\bf 520} 66-68  (2015). 

\bibitem{Barret2001} M. D. Barrett, J. A.Sauer,  and M. S. Chapman, Phys. Rev. Lett. {\bf 87} 010404 (2001).

\bibitem{Clement2009}  J.-F.  Cl\'ement, J.-P. Brantut, M. Robert-de-Saint-Vincent, R. Nyman, A. Aspect, T. Bourdel,  and P. Bouyer, Phys. Rev. A {\bf 79} 061406 (2009).

\bibitem{Lin2009} Y.-J. Lin, A. R. Perry, R. L. Compton, I. B.  Spielman, and J. V. Porto, Phys. Rev. A {\bf 79} 063631 (2009).

\bibitem{Bardou1992} F. Bardou, O.  Emile, J.-M. Courty, C. I. Westbrook, and A. Aspect, Europhys. Lett. {\bf 20} 681 (1992).

\bibitem{Mastwijk1998} H. C. Mastwijk, J. W.  Thomsen, P. van der Straten,  and A.  Niehaus, Phys. Rev. Lett {\bf 80} 25 (1998).

\bibitem{Kumakura1999} M. Kumakura,  and N. Morita,  Phys. Rev. Lett {\bf 82} 2848 (1999).

\bibitem{Browaeys2000} A. Browaeys, J. Poupard, A. Robert, S.  Nowak, W. Rooijakkers, E. Arimondo, L. Marcassa, D. Boiron, C. I. Westbrook,  and A. Aspect, Eur. Phys. J. D {\bf 8} 199-203 (2000).

\bibitem{Majorana1932} E. Majorana, Il Nuovo Cimento {\bf 9} 43 (1932).

\bibitem{Robert2001} A. Robert, O. Sirjean, A. Browaeys, J. Poupard, S. Nowak, D. Boiron, C. I. Westbrook, and A. Aspect., Science {\bf 292} 461 (2001)..

\bibitem{Santos2001} F. Pereira Dos Santos, J. Léonard, J. Wang, C. J. Barrelet, F. Perales, E. Rasel, C. S. Unnikrishnan, M. Leduc, and C. Cohen-Tannoudji,  Phys. Rev. Lett. {\bf 86} 3459 (2001).

\bibitem{Tychkov2006} A. S. Tychkov, T. Jeltes, J. M. McNamara, P. J. J. Tol, N. Herschbach, W. Hogervorst, and W. Vassen, W., Phys. Rev. A {\bf 73} 31603(R) (2006).

\bibitem{Dall2006} R. G. Dall and A. G. Truscott, Optics Comm. {\bf 270} 255-261 (2006).

\bibitem{Doret2009} S. C. Doret, C. B. Connolly, W. Ketterle, and J. M. Doyle, Phys. Rev. Lett. {\bf 103} 103005 (2009).

\bibitem{Keller2014} M .Keller, M. Kotyrba, F. Leupold, M. Singh, M. Ebner,  and A. Zeilinger, Phys. Rev. A {\bf 90} 063607 (2014).

\bibitem{Chang2014} R. Chang, A. L. Hoendervanger, Q. Bouton, Y. Fang, T. Klafka, K. Audo, A. Aspect, C. I. Westbrook,  and D. Cl\'ement, Phys. Rev. A {\bf 90} 063707 (2014).

\bibitem{Fluo} Fluorescence imaging with an InGaAs camera (Xeva-1.7-320 from Xenics) is used to characterize the atom cloud.

\bibitem{Lawal1995} J. Lawall, S. Kulin, B. Saubamea, N. Bigelow, M. Leduc, and C. Cohen-Tannoudji, Phys. Rev. Lett. {\bf 75} 4194 (1995).

\bibitem{Shlyapnikov1994} G. V. Shlyapnikov, J. T. M. Walraven,  U. M. Rahmanov,  and M. W. Reynolds, Phys. Rev. Lett. {\bf 73} 3247 (1994).

\bibitem{Heo2011} M.-S. Heo, J.-Y. Choi,   and Y.-I. Shin, Phys. Rev. A {\bf 83} 013622 (2011).

\bibitem{Petrich1995} W. Petrich, M. H. Anderson, J. R.  Ensher, and E. A. Cornell, Phys. Rev. Lett {\bf 74} 17 (1995).

\bibitem{Dubessy2012} R. Dubessy, K. Merloti, L. Longchambon, P.-E.  Pottie, T. Liennard, A. Perrin, V. Lorent, and H. Perrin, Phys. Rev. A  {\bf 85} 013643 (2012).

\bibitem{Browaeys2001} A. Browaeys, A. Robert, O. Sirjean, J. Poupard, S. Nowak, D. Boiron, C. I. Westbrook, and A. Aspect, Phys. Rev. A {\bf 64} 034703 (2001).

\bibitem{Ketterle1996} W. Ketterle,  and  N. J. van Druten, Adv. Atom. Mol. Opt. Phys. {\bf 37} 181-236 (1996).

\bibitem{Partridge2010} G. B. Partridge, J.-C.  Jaskula, M. Bonneau, D. Boiron, and C. I. Westbrook, Phys. Rev. A {\bf 81} 053631 (2010).

\bibitem{Stamper-Kurn1998} D. M. Stamper-Kurn, H.-J. Miesner, A. P. Chikkatur, S. Inouye, J. Stenger, and W. Ketterle, Phys. Rev. Lett. {\bf 81} 2194 (1998).

\bibitem{Hung2008} C.-L. Hung, X. Zhang, N. Gemelke, and C. Chin, Phy. Rev. A {\bf 78} 011604(R) (2008).

\bibitem{Dalfovo1999} F. Dalfovo, S. Giorgini, L. P. Pitaevskii,  and S. Stringari, Rev. Mod. Phys. {\bf 71} 463 (1999).


\bibitem{Bloch2008} I. Bloch, J. Dalibard and W. Zwerger, Rev. Mod. Phys. {\bf 80} 885-964 (2008).

\end{thebibliography}

\end{document}